\documentclass[aps,pra, superscriptaddress,reprint]{revtex4-1}

\usepackage{amsmath}
\usepackage{graphicx}
\usepackage{sistyle}
\usepackage{todonotes}
\usepackage{gensymb}
\usepackage{bbold}
\usepackage{bm}
\usepackage{natbib}
\usepackage{amsfonts}
\usepackage{dsfont}
\usepackage{empheq}
\usepackage{amssymb}
\usepackage{latexsym}
\usepackage{fancyhdr}
\usepackage[bbgreekl]{mathbbol}
\usepackage{epsfig}
\usepackage{color}

\newcommand{\sket}[1]{{\ensuremath{\lvert#1\rangle}}}
\newcommand{\lket}[1]{{\ensuremath{\left\lvert#1\right\rangle}}}
\newcommand{\ket}[1]{\if@display\lket{#1}\else\sket{#1}\fi}

\newcommand{\sbra}[1]{{\ensuremath{\langle#1\rvert}}}
\newcommand{\lbra}[1]{{\ensuremath{\left\langle#1\right\rvert}}}
\newcommand{\bra}[1]{\if@display\lbra{#1}\else\sbra{#1}\fi}

\newcommand{\sbraket}[2]{{\ensuremath{\langle#1\rvert#2\rangle}}}
\newcommand{\lbraket}[2]{{\ensuremath{\left\langle#1\!\left\rvert\vphantom{#1}#2\right.\!\right\rangle}}}
\newcommand{\braket}[2]{\if@display\lbraket{#1}{#2}\else\sbraket{#1}{#2}\fi}

\newcommand{\sketbra}[2]{{\ensuremath{\lvert #1\rangle\!\langle #2\rvert}}}
\newcommand{\lketbra}[2]{{\ensuremath{\left\lvert #1\right\rangle\!\!\left\langle #2\right\rvert}}}
\newcommand{\ketbra}[2]{\if@display\lketbra{#1}{#2}\else\sketbra{#1}{#2}\fi}

\newcommand{\proj}[1]{\ketbra{#1}{#1}}

\begin{document}

\title{Resource-efficient measurement device independent entanglement witness}

\date{\today}

\author{E.~Verbanis}
\affiliation{Group of Applied Physics, University of Geneva, Switzerland}
\author{A.~Martin}
\affiliation{Group of Applied Physics, University of Geneva, Switzerland}
\author{D.~Rosset}
\affiliation{Group of Applied Physics, University of Geneva, Switzerland}
\author{C.~C.~W.~Lim}
\affiliation{Group of Applied Physics, University of Geneva, Switzerland}
\affiliation{Oak Ridge National Laboratory, Oak Ridge, TN, United States}
\author{R.~T.~Thew}
\email{robert.thew@unige.ch}
\affiliation{Group of Applied Physics, University of Geneva, Switzerland}
\author{H.~Zbinden}
\affiliation{Group of Applied Physics, University of Geneva, Switzerland}

\date{\today}

\begin{abstract} 
Imperfections in experimental measurement schemes can lead to falsely identifying, or over estimating, entanglement in a quantum system. A recent solution to this is to define schemes that are robust to measurement imperfections - measurement device independent entanglement witness (MDI-EW). Here we introduce a novel approach for MDI-EW, which significantly reduces the experimental complexity and is applicable to a wide range of physical systems. The scheme requires no prior description of the state, is detection loop-hole free, robust to classical communication, and works for all entangled qubit states. We focus on photonic entanglement, experimentally generating and testing bipartite Werner states, varying the entanglement from the maximally entangled Bell state, past the bound for nonlocal states and down to the separable bound of 1/3. We witness entanglement down to an entangled state fraction close to 0.4. These results could be of particular interest for device independent quantum random number generation.
\end{abstract}
\maketitle

Entanglement is one of the quintessential characteristics of quantum physics and, importantly, is a crucial resource for quantum technologies~\cite{Walmsley2015}. While entanglement exists in many forms, here we focus on characterizing and quantifying entanglement in photonic systems. In this case, the most commonly used approach for generating entanglement is based on non-linear interactions, like spontaneous parametric down-conversion (SPDC)~\cite{Tanzilli2012,Spring2013,*Bruno2014}, which are commonly used in many quantum communication task~\cite{Gisin2007}.

In practice, there are three main approaches to characterize entangled states: Quantum state tomography (QST)~\cite{James2001,*Thew2002}, Entanglement Witnesses~\cite{Horodecki1996}, and Bell inequalities~\cite{Bell1964,*Genovese2005,*Brunner2014}. QST, uses a set of local measurements that are made on multiple copies of the unknown two-qubit state. This leads to an estimated density matrix $\hat{\rho}$ from which its fidelity with some target state, or the degree of entanglement, can be computed. However, this approach is prone to experimental errors, which can lead to non-physical states being reconstructed~\cite{DAriano2003}. Various techniques, such as maximum-likelihood or least-squares optimization are used to make the states physical, but this can also lead to the degree of entanglement being over-estimated~\cite{Schwemmer2015}. 

If one is only interested in certifying that the source generates entangled states, then an entanglement witness can be used~\cite{Guhne2009a}. However, if there are errors in the implementation of the measurements, for any of these approaches, then one cannot faithfully witness entanglement~\cite{Guhne2009a}. A way to characterise entanglement in a Device Independent manner is to rely on loophole-free violation of a Bell inequality~\cite{Acin2006,*Skwara2007}. This approach requires a high detection efficiency to close the detection loophole. Furthermore, it can only detect the entanglement of non-local states. 

A novel solution to overcome these problems was recently proposed by Branciard {\it et al.}~\cite{Branciard2013}, whereby instead of using classical inputs to perform a Bell test, these are replaced by quantum states~\footnote{This approach arose out of the work on semiquantum games by Buscemi~\cite{Buscemi2012}}. This approach also has a connection with measurement device independent quantum key distribution (QKD) protocols~\cite{Lo2012,*Xu2013} in the sense that a Bell state measurement is employed between the state under consideration and the auxiliary quantum input states. For this reason the resulting measurement device independent entanglement witness (MDI-EW) is faithful and does not necessitate closing the detection and locality loopholes~\cite{Rosset2013}. This is due to the fact that classical communication between Alice and Bob does not increase the violation of the MDI-EW.

\begin{figure}[h]
\includegraphics[width=0.45\textwidth]{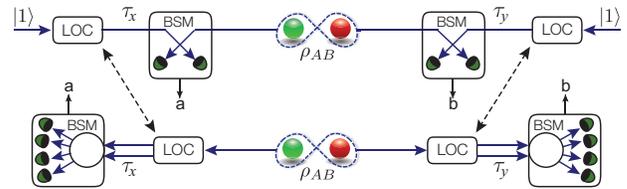}
\caption{Measurement Device Independent Entanglement Witness concept. Top: An entangled state is probed with locally prepared quantum inputs prepared by trusted Linear Optic Circuits (LOC) and the (2 possible) results of the Bell state measurements (BSM) are used to calculate the witness. Bottom: The simplified scheme uses trusted quantum states encoded on an extra degree of freedom of the initial entangled state using LOC and the (4 possible) BSM outcomes are used to reconstruct the witness.}
\label{fig_xconcept}
\end{figure}

Two photonic MDI-EW experiments have been performed to characterize bipartite entangled states in set-ups involving six photons~\cite{Xu2014,*Nawareg2015}. The concept is shown in the top of \figurename{~\ref{fig_xconcept}}. While these show the validity of MDI-EW, they are extremely demanding experiments. 

In this Letter we introduce and experimentally demonstrate a variation of the MDI-EW protocol, which greatly reduces the experimental overhead using concepts recently introduced for detector device independent QKD~\cite{Lim2014}. As presented  in \figurename{~\ref{fig_xconcept}} (bottom), our approach allows Alice and Bob to encode the input qubit states directly in an extra degree of freedom, thus removing the need of the ancillae photons. This has several advantages, namely, we reduce this from an experiment requiring six, identical and factorable~\cite{Spring2013,*Bruno2014}, photons, to only two, and we can also now detect all four Bell states. Firstly, we introduce a slightly different approach to construct a more general MDI-EW.

As shown in \figurename{~\ref{fig_xconcept}}, Alice and Bob share a quantum state $\rho_\text{AB}$. At each run of the experiment, Alice (Bob) prepares an input state $\tau_x$ ($\tau_y$), selected at random from the set $\{ \tau_1 ... \tau_m \}$ and each makes a joint measurement with part of the shared state $\rho_\text{AB}$; note that the indices $x,y$ of the selected states are recorded by Alice and Bob, and not provided to the devices. 

Describing the measurement of Alice (Bob) by the POVM $\{A_a\}$ ($\{B_b\}$), the following correlations are observed:
\begin{eqnarray} \label{Eq:Pabxy:Full}
P(a b| \tau_x, \tau_y) = \text{tr} \left [ \left (A_a \otimes B_b \right ) \left ( \tau_x \otimes \rho_\text{AB} \otimes \tau_y \right ) \right ].
\end{eqnarray}
When Alice and Bob share a separable state $\rho_\text{AB} = \sum_k \rho^{\text{A}}_k \otimes
\rho^{\text{B}}_k$, with $\rho^{\text{A}}_k, \rho^{\text{B}}_k \geqslant 0$,
their correlations are given by:
\begin{equation}
  \label{Eq:Pabxy:SEP} P_{\text{SEP}} (a b | \tau_x \tau_y) = \sum_k \mathrm{tr} \left[
  \mathrm{A}_a \left( \tau_x \otimes \rho_k^{\text{A}} \right) \right] \mathrm{tr}
  \left[ \mathrm{B}_b \left( \rho_k^{\text{B}} \otimes \tau_y \right) \right] .
\end{equation}
A MDI-EW is defined as:
\begin{eqnarray} \label{eq:MDIEW:calc}
W = \sum_{abxy} \beta_{abxy} \cdot P(a b| \tau_x, \tau_y),
\end{eqnarray}
with the following properties:
\begin{itemize}
\item $W < 0$ for a particular entangled state $\rho_\text{AB}$ and specific measurements $\{A_a\}$, $\{B_b\}$;
\item $W \ge 0$ when $\rho_\text{AB}$ is separable, for all possible measurements $\{A_a\}$, $\{B_b\}$ (or, more generally, when Alice and Bob share any classical resource, see below).
\end{itemize}
The MDI-EW is thus characterized by a set of input states $\{ \tau_{x, y} \}$ and real coefficients $\beta_{abxy}$, and its violation certifies the presence of entanglement in $\rho_\text{AB}$, without trusting the measurement devices.

Any entangled state can be detected by a suitable MDI-EW, as described in~\cite{Buscemi2012, Branciard2013, Rosset2013}; while the MDI-EW construction does not assume that particular measurements are implemented by Alice and Bob, it is violated only when the shared state is close to the one used in the MDI-EW construction, and when the measurements are close to the Bell state measurements in prescribed bases. 

Conversely, let us consider a MDI-EW-like scenario, where correlations $P(ab|\tau_x,\tau_y)$ are observed using a well-characterized set of input states $\{\tau_{x,y}\}$, without however having the relevant witness coefficients $\beta_{abxy}$. We show below how to construct a value $W'$ having the same properties as in Eq.~\eqref{eq:MDIEW:calc} ($W' \ge 0$ for separable resources), and then, in a second step, how to create a MDI-EW tailored for the setup considered. We sketch below the construction to be expanded on in future work~\cite{Rosset_next}.

We first observe that a compact description of the experimental setup, including the state $\rho_\text{AB}$ as well as the measurements $\{ \mathrm{A}_a \}$ and $\{ \mathrm{B}_b \}$, is provided by the joint POVM $\{ \Pi_{a b} \}$ acting on the input state $\tau_x \otimes \tau_y$:
\begin{equation} \label{Eq:Pabxy:Pi}
  P (a b | \tau_x, \tau_y) = \mathrm{tr} [\Pi_{a b} (\tau_x \otimes \tau_y)].
\end{equation}
This description is slightly more general than Eq.~(\ref{Eq:Pabxy:Full}); for example, it allows classical communication between Alice and Bob's devices. However, when Alice and Bob share a separable state $\rho_\text{AB} = \rho_\text{AB}^{\text{SEP}}$, the POVM elements $\Pi_{a b} = \Pi^{\text{SEP}}_{a b}$ are separable:
\begin{equation}
  \label{Eq:PiSEP} \Pi^{\text{SEP}}_{a b} = \sum_k \Pi^{\text{A}}_{a,
  k} \otimes \Pi_{b, k}^{\text{B}},
\end{equation}
decomposed over $\Pi_{a, k}^{\text{A}}, \Pi^{\text{B}}_{b, k} \geqslant 0$.
This can be seen from Eq.~(\ref{Eq:Pabxy:SEP}), and in general, we have
$\Pi_{a b} = \Pi_{a b}^{\text{SEP}}$ when Alice and Bob share classical
resources~\cite{Chitambar2014}. The partial transpose of (\ref{Eq:PiSEP}), $ \left( \Pi_{a b}^{\text{SEP}} \right)^{\top_{\text{A}}} = \sum_k (
  \Pi^{\text{A}}_{a, k} )^{\top} \otimes \Pi_{b, k}^{\text{B}}
  \geqslant 0 $ is nonnegative. Conversely, nonseparable $\Pi_{a b}$ can have partial transposes
 $(\Pi_{a b})^{\top_{\text{A}}}$ with negative eigenvalues --- and for qubits, nonseparable operators always have negative partial transposes~\cite{Horodecki1996}. We decompose $(\Pi_{ab})^{\top_{\text{A}}}$ in parts with positive and negative eigenvalues~\cite{Moroder2013}:
\begin{equation}\label{Eq:Pi:Decomp}
  (\Pi_{a b})^{\top_{\text{A}}} = \sigma_{a b}^+ - \sigma_{a b}^-, \quad W' = -\sum_{ab} \min \mathrm{tr} [\sigma_{a b}^-] ,
\end{equation}
with $\sigma_{a b}^{\pm} \geqslant 0$. Clearly, when $\rho_\text{AB}$ is separable (or Alice and Bob share classical resources), all $\Pi_{a b}$ are separable, and the minimum is obtained for $\sigma_{a
b}^+ = (\Pi_{a b})^{\top_{\text{A}}}$, $\sigma_{a b}^- = 0$; thus $W' = 0$. Conversely, $W' < 0$ certifies the presence of entanglement in $\rho_\text{AB}$. Thus, the value $W'$ satisfies the properties outlined after Eq.~\eqref{eq:MDIEW:calc}; it can be easily computed using a semidefinite solver~\cite{Stu:99}, as equations~\eqref{Eq:Pabxy:Pi} and~\eqref{Eq:Pi:Decomp} define a semidefinite program.

We show in~\cite{SuppMat} how to extract MDI-EW coefficients from the computation of $W'$, recovering the familiar form of Eq.~\eqref{eq:MDIEW:calc}. This MDI-EW will be optimal for the current setup with probabilities $P(ab|\tau_x,\tau_y)$, but can nevertheless be applied to other setups as a valid MDI-EW as long as the same set of input states $\{ \tau_{x,y} \}$ is employed.

\begin{figure*}[]
\includegraphics[width=0.9\textwidth]{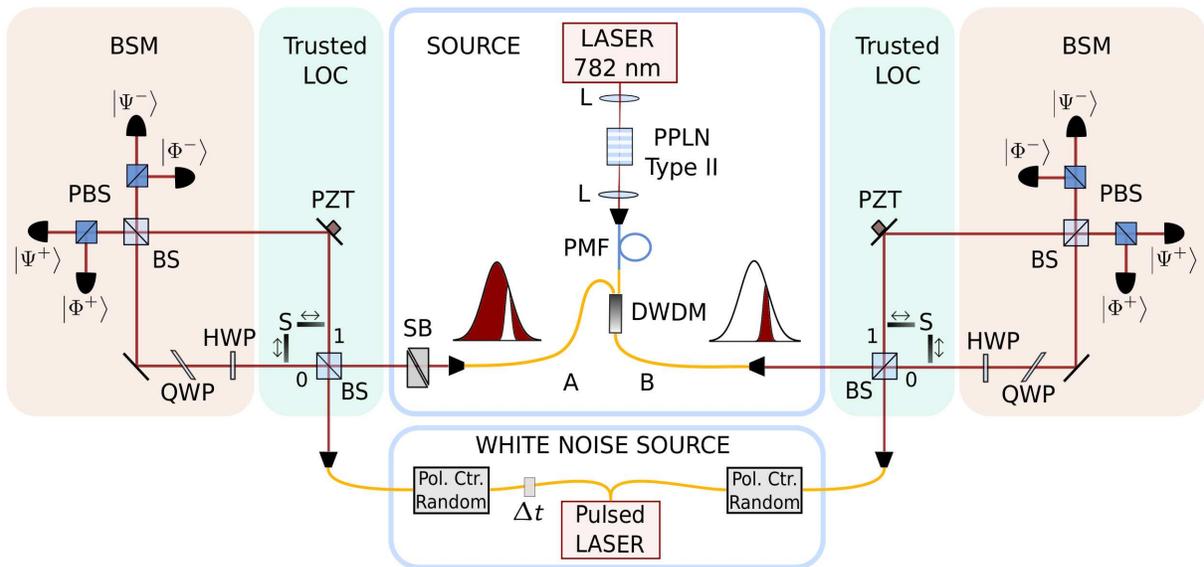}
\caption{Experimental set-up. Polarisation entangled photon pairs are produced by first pumping a type-II periodically poled lithium niobate (PPLN) non-linear crystal with a continuous laser, then by deterministically separating the degenerate photons using a single channel dense wavelength division multiplexer (DWDM).  The input qubits $\tau_x$ and $\tau_y$ are encoded directly onto the path of the corresponding photon via a trusted linear optical circuit (LOC). Alice and Bob perform Bell state measurements  (BSM) using a half-wave plate (HWP) in the lower arm of the interferometers, and polarizing beam-splitters (PBS) on each output arms followed by single photon detectors. L: lens; PMF: polarisation maintaining fiber; SB: Soleil-Babinet; BS: beam-splitter; S: shutter; QWP: quarter-wave plate; PZT: piezoelectric transducer.}
\label{fig_schema}
\end{figure*}

We now consider the problem of low detection efficiencies and losses. In our setup, we regroup all events where non-detections occur (on either side) under an additional outcome $\emptyset$, such that:
\begin{equation}\label{Eq:Pabxy:Norm}
P(\emptyset|\tau_x, \tau_y) + \sum_{abxy} P(ab|\tau_x, \tau_y) = 1 \text{ for all } x,y.
\end{equation}
Let $P_\eta(ab|\tau_x, \tau_y) = \eta P(ab|\tau_x, \tau_y)$, with $P_\eta(\emptyset|\tau_x, \tau_y)$ satisfying~Eq.~\eqref{Eq:Pabxy:Norm}, be the correlations observed according to some efficiency $\eta > 0$. Then, if the original $P$ violates the MDI-EW with $W < 0$, then $P_\eta$ has $W_\eta = \eta W < 0$, as already described in~\cite{Branciard2013}.

In experiments involving CW-based SPDC sources  the fraction of non-detection events is unknown due to the random emission and detection times.
In this case, we apply the following estimation technique for $P(ab|\tau_x, \tau_y)$. Let $N(abxy)$ be the number of observed events for outputs $a,b$ and input states $\tau_x, \tau_y$, and $N(\emptyset x y)$ the (unknown) number of non-detections for inputs $\tau_x, \tau_y$. When the experiment is run for the same amount of time for all pairs $(x,y)$, with constant efficiency, we have $N(x y) = N(\emptyset x y) + \sum_{ab} N(abxy) = N^*$, a constant. As observed, the MDI-EW construction is insensitive to any rescaling $P \rightarrow \eta P$, and $N^*$ can be chosen arbitrarily. For the MDI-EW test, we use the following values:
\begin{equation} \label{eq:Pabxy:recons}
P(ab|\tau_x, \tau_y) = \frac{N(abxy)}{N^*}, \quad N^* = \max_{xy} \sum_{ab} N(abxy),
\end{equation}
such that $0 \le P(ab|\tau_x, \tau_y) \le 1$.

MDI-EW is robust against the locality loophole (classical communication between the device is allowed), and against the detection loophole. By construction, any strategy based on separable ressources using non-detections is already included in the POVMs of Eq.\eqref{Eq:PiSEP}~\footnote{These properties are already discussed separately in Refs.~\cite{Branciard2013,Rosset2013} and will be reviewed in Ref.\cite{Rosset_next}}. 

The experimental setup is shown in \figurename{\ref{fig_schema}}. The entangled photon pairs centered around 1564\,nm are generated by a SPDC process in a type-II periodically poled lithium niobate (PPLN) crystal, pumped by a CW laser at 782\,nm with a power of 50\,mW. To compensate the temporal walk-off induced by the crystal birefringence between the two orthogonally polarized photons, the photon pairs are directly coupled in a 1.44 m long polarization maintaining fiber (PMF) ~\cite{Kaiser2012,Bruno2013}. The photons are deterministically separated using a 100 GHz single channel dense wavelength division multiplexer (DWDM) slightly detuned from the central wavelength emission. The energy conservation associated with the CW pump laser introduces strong wavelength correlations such that a polarization entangled state of the following form is produced: $ \ket{\Psi^+}_{AB} = \frac{1}{\sqrt{2}}\left[\ket H_A \ket V_B +  \ket V_A \ket H_B\right],$
where $A$ and $B$ represent Alice and Bob, respectively. A $g^{(2)}(0)$ on the order of $10^{-3}$ ensures that the double pair contribution is negligible. The relative phase is adjusted via the Soleil-Babinet (SB) placed on Alice's side, to generate the $ \ket{\Psi^+}_{AB}$ polarization Bell state. 

The inputs qubits $\tau_x$ and $\tau_y$ are encoded onto the optical path degree of freedom via two 50/50 beamsplitters (BS) placed on each side (see \figurename{\ref{fig_schema}). The qubit states are of the form  $\ket\tau_j = \left[\ket 0+e^{i\varphi_{j}}\ket1\right]/\sqrt{2}$, where we assign  $\ket 0$ and $\ket 1$ to the lower and upper arm, respectively. To set the phases $\varphi_{j}$ a piezoelectric transducer (PZT) is mounted on a mirror on the upper arm of the interferometer. The two other states $\ket 0$ and $\ket 1$ are obtained by blocking the appropriate arm of the interferometer using automated shutters (S).  It can be noted that to compensate the birefringence introduced by the optical elements in the interferometer a tilted quarter-wave plate (QWP) is added in the lower arms. 
 
 A complete Bell state measurement (BSM) is performed by first transforming $\ket{H}$ to $\ket{V}$ and  $\ket{V}$ to  $\ket{H}$ on the lower arm using a half-wave plate (HWP), then by recombining the two arms on a 50/50 BS, and finally by projecting in the \{\ket{H},\ket{V}\} basis using polarizing beam-splitters (PBS) on both output arm. Each output of the PBS corresponds to one of the following Bell states (see \figurename{~\ref{fig_schema}}):
\begin{equation}
\begin{split}
\ket{\Psi^\pm}=\frac{1}{\sqrt{2}}\left(\ket{H}\ket{1}\pm\ket{V}\ket{0}\right), \\
\ket{\Phi^\pm}=\frac{1}{\sqrt{2}}\left(\ket{H}\ket{0}\pm\ket{V}\ket{1}\right).
\end{split}
\end{equation}
To perform the measurement, the twofold coincidences between the four single photon detectors (ID220 with an efficiency of 20\% and around 1kHz of dark-count) of Alice and Bob are recorded via a time-to-digital converter (TDC).  

To produce a Werner state of the form: 
\begin{equation}\label{wernerstate}
\rho_\text{AB}  = \lambda  \proj{\Psi^+}_\text{AB} + (1-\lambda)\mathbb{1}_4
\end{equation}
an additional pulsed, telecom-wavelength, laser is injected inside the interferometers together with the photon pairs. The relative arrival time of the pulses in the interferometers of Alice and Bob is set to observe coincidence peaks in the same temporal windows as the photon pairs. Two electronic adjustable polarisation controllers driven by two uncorrelated random sequences are employed to obtain an unpolarised noise. Moreover, to decrease the Bell state weight $\lambda$, the repetition rate of the pulsed laser is simply increased. 

The set of input states employed to certify the entanglement of the Werner states down to $\lambda=1/3$ is $\{\tau_{x,y}\}=\{\ket{0}\pm\ket{1}; \ket{0}\pm i\ket{1};  \ket{0};  \ket{1}\}$.  The coincidence counts from the BSMs are recorded for all thirty-six  pairs of input states. Without additional noise, the average coincidence rate is about 16 counts per second for each output, corresponding to a total detection efficiency around $3\%$. The integration time is set to 10 seconds for each input pair such that, using automatic control of PZTs and shutters, one complete measurement lasts about 6 minutes. 

For each Bell state fraction $\lambda$, the number of observed events $N_\lambda(abxy)$ is collected, from which we reconstruct the probability distribution $P_\lambda(ab|\tau_x,\tau_y)$ according to Eq.~\eqref{eq:Pabxy:recons}. However, due to finite statistics and noise, the distribution $P_\lambda$ does not satisfy Eq.~\eqref{Eq:Pabxy:Pi} for $\Pi_{ab} \ge 0$; thus $W'$ cannot be computed directly from $P_\lambda$ as the semidefinite program is infeasible. The slight inconsistencies in $P_\lambda(ab|\tau_x,\tau_y)$ are corrected by looking for the closest regularized distribution $\overline{P}_\lambda$ satisfying Eq.~\eqref{Eq:Pabxy:Pi} with $\Pi_{ab} \ge 0$ --- we use the Euclidean distance so that the computation is another semidefinite program. Then, using the method of~\cite{SuppMat}, we construct a MDI-EW with the coefficients $\beta^\lambda_{abxy}$ tailored for the probabilities $\overline{P}_\lambda$. However, we compute the witness value $W_\lambda = \sum_{abxy} \beta^\lambda_{abxy} P_\lambda(ab|\tau_x,\tau_y)$ using the original distribution $P_\lambda$, thus avoiding the introduction of a bias in the estimation of $W_\lambda$~\cite{Schwemmer2015}. The computed witness values are plotted in \figurename{\ref{fig_result1}} (square data points), starting from $\lambda=0.94$. Note, $\lambda<1$ due to the intrinsic noise of the detectors. We observe that the witness value saturates at $W = 0$ when entanglement cannot be certified; this comes from the minimization present in Eq.~\eqref{Eq:Pi:Decomp}

While convenient, this first approach has the disadvantage of using every dataset twice, first to construct a MDI-EW, then to estimate the MDI-EW value. A more robust approach is to compute the MDI-EW coefficients once for some $\lambda$, and then apply the MDI-EW to the others. We thus use $\hat{\beta}_{abxy} = \beta^{\lambda=0.94}_{abxy}$ to compute the values $\hat{W}_\lambda = \sum_{abxy} \hat{\beta}_{abxy} P_\lambda(ab|\tau_x,\tau_y)$, also plotted in \figurename{~\ref{fig_result1}} (circle data points). By linearity of Eq.\eqref{wernerstate}, the witness value $\hat{W}_\lambda$ becomes positive for separable Werner states. For $\lambda > 0.4$, this second robust method performs slightly worse, but avoids overadapting the MDI-EW to the noise present in the correlations. 

In both approaches, the entanglement of the Werner states could not be certified all the way down to the separability limit $\lambda=1/3$. The reason for this is that the values of the witness $W$ are limited by the imperfections in the BSMs and the residual birefringence inside the interferometers. They both induce small phase shifts between the outputs of the BSMs and hence, effectively reduce the value of the witness. The value $W$ in \figurename{\ref{fig_result1}} can be related to a lower bound on the amount of entanglement present in the Werner state $\rho_\text{AB}$, as quantified by the negativity~\cite{Rosset_next}. 

\begin{figure}[tp]
\includegraphics[width=0.5\textwidth]{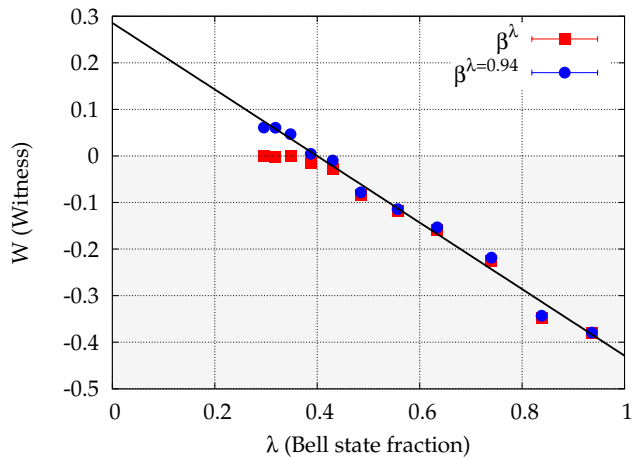}
\caption{Witness values for two-qubit Werner states with different weights $\lambda$. The data points $\beta^\lambda$ and  $\beta^{\lambda=0.94}$ correspond to the results obtained when $\beta$ are calculated for each points and only for the first point, respectively. The uncertainty associated with each point has been calculated using Monte-Carlo algorithm and a Poissonian noise on the detection rate.}
\label{fig_result1}
\end{figure}

\paragraph{Conclusion} --
We have proposed and demonstrated a novel and practical measurement device independent entanglement witness protocol, certifying entanglement for a family of polarization Werner states down to a Bell state weight close to 0.4. Our approach replaces the need for additional single photons to encode the input qubits by encoding these directly on an extra degree of freedom of the entangled photons. This has the advantage that even entangled states encoded on photon pairs that are non-factorable in the spectral domain can be characterised. Moreover, all four Bell state measurement outcomes can be used to reconstruct the witness, which can be constructed directly from the measured output probabilities. Given that this MDI-EW is robust against the locality and detection loopholes, it opens up the interesting question as to whether a similar approach could be exploited to realize a more practical device independent quantum random number generator, as it would overcome the problems of needing space-like separated inputs and having low detection efficiencies.

\begin{acknowledgments}
The authors would like to thank Natalia Bruno, Flavien Hirsch, Nicolas Brunner, and Nicolas Gisin for discussions. This work was supported by the Swiss national sciences foundation project 200021-159592.
\end{acknowledgments}

\bibliography{MDIEW}

\appendix

\section{Appendix A: Recovering MDI-EWs from semidefinite program solutions}
\label{AppendixA}
We collect together Eqs.~\eqref{Eq:Pabxy:Pi} and \eqref{Eq:Pi:Decomp} in the following optimization problem:
\begin{empheq}[box=\fbox]{align}
-W' = \min_{\Pi_{ab}, \sigma_{ab}^\pm}  \sum_{ab} \mathrm{tr} [\sigma_{a b}^-] & \text{ s.t.} \nonumber \\
\Pi_{ab}, \hspace{0.5em} \sigma_{ab}^\pm & \geqslant 0 \nonumber \\
\sigma_{ab}^+ - \sigma_{ab}^- - (\Pi_{ab})^{\top_{\mathrm{A}}} & = 0 \nonumber \\
 \mathrm{tr} [\Pi_{ab} (\tau_x \otimes \tau_y)] & =  P (a b | \tau_x \tau_y) \label{Eq:Primal}
\end{empheq}

This semidefinite program is readily solved by standard packages~\cite{Stu:99} for given $\{ \tau_{x, y} \}$, using $P (a b | x y)$ obtained experimentally; a result $W' > 0$ certifies the presence of entanglement.

The dual problem is given by:
\begin{empheq}[box=\fbox]{align}
D = \max_{\gamma_{abxy}, Y_{ab}}  \sum_{abxy} \gamma_{abxy} P(ab|\tau_x, \tau_y) & \text{ s.t.} \nonumber \\
Y_{ab} & \geqslant 0 \nonumber \\
\mathbb{1} - Y_{ab} & \geqslant 0 \nonumber \\
Y_{ab}^{\top_{\mathrm{A}}} - \sum_{abxy} \gamma_{abxy} \left ( \tau_x \otimes \tau_y \right ) & \geqslant 0 \label{Eq:Dual}
\end{empheq}

From weak duality~\cite{Boyd2004}, the solution of the dual~\eqref{Eq:Dual} provides a bound on the solution of the primal~\eqref{Eq:Primal}: $ D \le -W' $, and a standard solver returns a solution of the primal and the dual problem. In particular, after solving the problem for particular $P(ab|\tau_x, \tau_y)$ and $\{ \tau_{x,y} \}$, the solution contains the coefficients $\beta_{abxy}$, that can be used as a MDI-EW for the set of inputs $\{ \tau_{x,y} \}$.

Indeed, only the objective function of the dual problem~\eqref{Eq:Dual} refers to the values $P(ab|\tau_x,\tau_y)$ and not the constraints; after solving~\eqref{Eq:Dual} for given $P(ab|\tau_x, \tau_y)$, the values $\gamma_{abxy}$ and $Y_{ab}$ are feasible for any $P'(ab|\tau_x, \tau_y)$. While $D' = \sum_{abxy} \gamma_{abxy} P'(ab|\tau_x, \tau_y)$ is not necessarily maximal --- indeed, the coefficients $\gamma_{abxy}$ were computed for $P(ab|\tau_x, \tau_y)$ --- the value $D'$ is a valid lower bound on the solution of the primal problem~\eqref{Eq:Primal} for $P'(ab|\tau_x, \tau_y)$. Thus, $D' > 0$ is a signature of entanglement, while, by construction, $D' \le 0$ for separable resources. To cater for the sign convention, we construct a MDI-EW using $\beta_{abxy} = - \gamma_{abxy}$.

\end{document}